\documentclass[12pt]{article}

\usepackage{amsmath,amssymb,array,amsfonts}
\usepackage[dvipdfmx]{graphicx}
\usepackage{comment,color}
\usepackage{ascmac}
\usepackage{cite}

\newcommand{\ba}{\begin{eqnarray}}
\newcommand{\ea}{\end{eqnarray}}

\allowdisplaybreaks

\def\ncm{\newcommand}
\def\M {{\rm M}}
\def\e {{\rm e}}

\def\rmd{{\rm d}}

\def\dis{\displaystyle}

\def\nt{\notag}

\ncm{\sls}[1]{{\ooalign{\hfil/\hfil\crcr$#1$}} }

\makeatletter
    
    \@addtoreset{equation}{section}
\makeatother

\begin{document}
\setlength{\baselineskip}{18pt}

\begin{titlepage}

\begin{flushright}
OCU-PHYS-514 \\
NITEP 40
\end{flushright}
\vspace{1.0cm}
\begin{center}
{\Large\bf 
Strong First Order Electroweak Phase Transition \\
\vspace*{3mm}
in Gauge-Higgs Unification at Finite Temperature
} 
\end{center}
\vspace{25mm}

\begin{center}
{\Large
Yuki Adachi 
and 
Nobuhito Maru$^{a,b}$
}
\end{center}
\vspace{1cm}
\centerline{{\it
Department of Sciences, Matsue College of Technology,
Matsue 690-8518, Japan}}
\vspace*{2mm}

\centerline{{\it
${}^{a}$Department of Mathematics and Physics, Osaka City University, Osaka 558-8585, Japan
}}
\vspace*{2mm}
\centerline{{\it
${}^{b}$Nambu Yoichiro Institute of Theoretical and Experimental Physics (NITEP), }}
\centerline{{\it
Osaka City University, Osaka 558-8585, Japan
}}
%
%
%
\vspace{2cm}
\centerline{\large\bf Abstract}
\vspace{0.5cm}
We analyze the electroweak phase transition at finite temperature in a model of gauge-Higgs unification 
 where the fermion mass hierarchy including top quark mass, a viable electroweak symmetry breaking 
 and an observed Higgs mass are successfully reproduced.
To study the phase transition, 
 we derive the general formula of the 1-loop effective potential at finite temperature 
 by using the $\zeta$ function regularization method.
It is remarkable that the functions determining the Kaluza-Klein mass spectrum have only to be necessary in calculations. 
This potential can be applicable to any higher dimensional theory in flat space 
 where one extra spatial dimension is compactified. 
Applying to our model of gauge-Higgs unification, 
 the strong first phase transition compatible with 125 GeV Higgs mass is found to happen.


\end{titlepage}

\clearpage
\section{Introduction}

Gauge-Higgs unification (GHU) \cite{GH, HIL} is one of the attractive scenarios 
 that solves the hierarchy problem without invoking supersymmetry, 
 where the Standard Model (SM) Higgs boson mass and its potential are calculable 
 thanks to the higher dimensional gauge symmetry \cite{HIL}.
These characteristic properties have been studied and verified in models 
 with various types of compactification at one-loop level \cite{ABQ}
 and at the two-loop level \cite{MY, HSY}. 
The calculability of other physical observables 
 have been also investigated \cite{LM, Maru, ALM}. 
 The flavor physics which is a very nontrivial in GHU has been studied in \cite{flavorGHU}.

In five dimensional (5D) GHU, 
  since Higgs potential at the tree level is forbidden by the gauge symmetry in higher dimensions,  
 but it is radiatively generated,  
 it is nontrivial to obtain a realistic electroweak symmetry breaking and the observed Higgs mass. 
In GHU, Higgs quartic coupling is provided by the gauge coupling squared and is 1-loop suppressed. 
Therefore, Higgs mass squared is likely to be light. 
In order to obtain an observed Higgs mass and a realistic electroweak symmetry breaking, 
 a very small Higgs vacuum expectation value (VEV) is required in GHU. 
It is well-known for getting small Higgs VEV that Higgs potential has to be generated 
 by various contributions from higher rank representations of the gauge group \cite{SSS, CCP}. 

As for the SM fermion masses, 
 embedding the SM fermions except for top quark into some massive bulk fermions, 
 Yukawa couplings can be obtained from the overlaps of zero mode functions of the gauge coupling. 
The fermion masses are easily reproduced by mild tuning of the bulk masses. 
Top quark should be embedded into massless bulk fermion to avoid a suppression, 
 and into a fermion with higher rank representation \cite{SSS, CCP}. 
 
In our previous paper \cite{Adachi:2018mby}, 
 we have proposed a new model with a greatly simplified fermion content, 
 where the fermion mass hierarchy including top quark mass, a successful electroweak symmetry breaking, 
 and 125 GeV Higgs mass are reproduced. 
The point of the model is that we have employed another mechanism 
 of generating Yukawa coupling for the third generation quarks. 
The third generation quarks are introduced as the brane-localized fermions not bulk fermions 
 and have couplings with bulk fermions through the mass term on the brane. 
Integrating out the bulk fermions lead to Yukawa coupling and we have succeeded in reproducing top quark mass. 
 
As a familiar application of the electroweak model to the finite temperature theory, 
 there exits an electroweak baryogenesis for the generation of baryon asymmetry.   
It is well known that an application of the SM to the electroweak baryogenesis does not work 
 because the 125 GeV Higgs mass is not compatible with the strong first order phase transition. 
Therefore, the application to the electroweak baryogenesis is well motivates to consider the physics beyond the SM. 

In a paper by one of the authors \cite{MT}, 
 an application of GHU to the electroweak phase transition at finite temperature has been considered. 
Although the models discussed in the paper were not realistic ones, 
 we have investigated whether strong first order electroweak phase transition takes place. 
The results were positive, namely, 
 the strong first order phase transition in GHU is relatively easy to happen 
 and a compatiblity with 125 GeV Higgs mass was suggested. 
The reason is that the higher dimensional gauge boson contribution to 
 the cubic term in Higgs potential at finite temperature is large. 
The application of a realistic GHU to the electroweak phase transition at finite temperature was found in \cite{PS}. 
Their result was that the phase transition becomes of first order, 
 but not strong enough for 125 GeV Higgs mass.   

In this paper, we investigate the phase transition at finite temperature 
 in our realistic GHU model mentioned above \cite{Adachi:2018mby}. 
In our model, it is very nontrivial to calculate the 1-loop effective potential at finite temperature 
 because of the brane localized fermions and their coupling to the bulk fermions. 
In such a case, the Kaluza-Klein (KK) mass spectrum cannot be exactly solved in general. 
Therefore, Poisson resummation technique familiar with the calculation of the 1-loop effective potential in higher dimensions 
 cannot be utilized. 
Instead, we employ the $\zeta$ function regularization method to calculate the 1-loop effective potential at finite temperature. 
The advantage of this method is that the functions determining the KK mass spectrum have only to be known, 
 but the KK mass spectrum themselves are not necessary.  
Along this line, we derive a general formula of 1-loop effective potential at finite temperature, 
 which is applicable to any higher dimensional models in flat space 
 where one extra spatial dimension is compactified. 
Applying this general potential to our model \cite{Adachi:2018mby}, 
 we analyze the electroweak phase transition at finite temperature 
 and find it to be strong first order compatible 125 GeV Higgs mass.

This paper is organized as follows. 
In section 2, we briefly describe our model. 
In section 3, we calculate the 1-loop effective potential at finite temperature 
 by exploiting $\zeta$ function regularization method and show the general formula.   
The electroweak phase transition at finite temperature is analyzed in section 4. 
Summary is given in section 5. 
In appendix, the calculation of the 1-loop effective potential at finite temperature is derived in detail.


\section{The model}
\label{section:model}
We begin with a brief review of the model proposed in our previous paper \cite{Adachi:2018mby}. 
The $SU(3)\otimes U(1)_X$ gauge theory in five-dimensional flat space-time is considered.
The fifth spatial extra dimension is compactified on an orbifold $S^1/Z_2$ with the radius $R$ of $S^1$.
The $U(1)_X$ gauge symmetry is introduced in order to realize the correct weak mixing angle $\theta_W$.

\begin{figure}
\begin{center}
\includegraphics{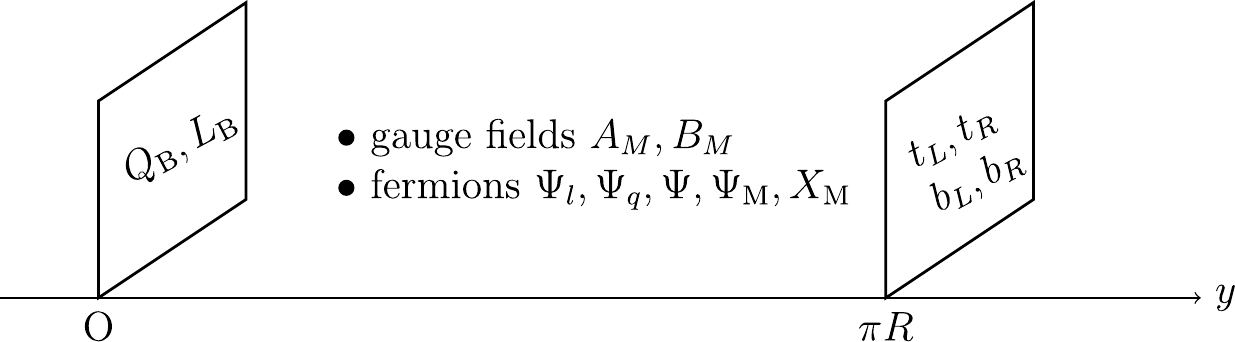}
\caption{Setup of the model.}
\label{fig:outline}
\end{center}
\end{figure}

The top ($t$) and the bottom ($b$) quarks in our setup 
 are brane-localized fermions 
 at the $y=\pi R$ brane.
The SM fermions other than the top and bottom quarks are embedded 
 in the bulk fermions $\Psi_l$ for leptons and $\Psi_q$ for quarks, 
 which are assigned to the fundamental representation $\bf 3$ of $SU(3)$.
They obtain a mass through the five-dimensional gauge interaction, 
 which is Yukawa interaction in the context of the gauge-Higgs unification scenario. 
Since the $t$ and $b$ quarks cannot interact directly with the Higgs boson ($A_y$) 
 being the fifth component of the gauge field in five dimensions,
 two extra bulk fermions $\Psi$ (referred to as messenger fermions) are introduced, 
 where one (the other) messenger fermion is embedded in the $\bf 3~(\bf 15)$ representation of $SU(3)$ 
  coupling to the bottom (top) quark on the $y=\pi R$ brane.
We also introduce a pair of fermions (referred to as mirror fermions)
$\Psi_\M$ and $X_\M$ in $\bf 15$ representation of $SU(3)$ 
 to realize the realistic electroweak symmetry breaking. 
Such fermions may be a possible candidate of the dark matter as pointed out in \cite{Maru:2017otg}.
The outline of the model is shown in the Figure \ref{fig:outline}.

As was pointed out in our previous paper, 
 this simple matter content can explain the fermion mass hierarchy including top quark mass 
 and the suitable electroweak symmetry breaking.
The reason for introducing fermions of higher dimensional representation such as $\bf 15$ 
 is to reproduce the Higgs mass. 
In GHU, Higgs mass is likely to be light because Higgs potential is generated by quantum corrections. 
In order to obtain 125GeV Higgs mass, the small VEV is required and 
 realized by utilizing their nature of high frequency of Higgs potential.   

To discuss the phase transition at finite temperature,
 we need the mass spectrum of the fields in our model
 because the non-zero KK modes contribute to the Higgs potential at 1-loop.
In general, the KK mass spectrum becomes very complicated in the presence of the brane-localized terms. 
It is therefore very hard to find the exact mass eigenvalues.
However, as will be explained in the next section, 
 if we employ the $\zeta$ function regularization method, 
 the mass eigenvalues are not required explicitly, 
 but only the conditions for the mass spectrum to be satisfied are necessary. 
Then, it allows us to compute the Higgs potential in detail.
These conditions are determined from the boundary conditions 
 such as the $Z_2$ symmetry and/or the periodic/anti-periodic boundary conditions.


\section{1-loop effective potential at finite temperature in $\zeta$ function renormalization method}

In this section, we discuss a general formula of 1-loop effective potential at finite temperature, 
 which can be applicable to any higher dimensional theory in flat space 
 where the mass spectrum cannot be found and one extra spatial dimension is compactified. 

First of all, we provide the formula to calculate the 1-loop effective potential at finite temperature.
A particle with the mass $m_n$ contributes to the 1-loop effective potential as
\begin{equation}
V
 =(-1)^F
 \frac{N_\text{DOF}}{\beta} \sum_{l=-\infty}^{\infty}\sum_{n=-\infty}^{\infty}
 \int\frac{\rmd^{D-1} p}{(2\pi)^{D-1}}\frac12 \ln(\vec p^2 +\omega^2 +m_n^2)
\end{equation}
where $\beta=T^{-1}$ is an inverse of temperature $T$. 
$\omega$ stands for the Matsubara frequency, 
 which is given by $\omega_B=\frac{2\pi}{\beta}l$ for the bosonic field and 
 $\omega_F=\frac{\pi}{\beta}(2l-1)$ for the fermionic field.
The $N_\text{DOF}$ stands for the degrees of freedom of the particles running in the loop. 
$F$ means a fermion number, $F=0 (1)$ for bosons (fermions). 
Note that the momentum integration covers $D-1$ dimensional momentum space
 because the time direction is compactified on a circle 
 in an imaginary time formalism of the finite temperature theory.
The Poisson resummation formula can be applied as usual for the simple form of KK mass spectrum such as $n/R$,
 but it is very difficult to do when the exact mass eigenvalues are not found. 
This is why we adopt the $\zeta$ function regularization method instead of the Poisson resummation formula.
The derivation of the above expression is described in detail in the Appendix.
The above 1-loop effective potential can be rewritten in terms of the $\zeta$ function regularization method as follows. 
\begin{align}
 \label{eq:EPinfnitetemperature}
V
 =& \nt
 -(-1)^F
 \frac{N_\text{DOF}}{2\beta} \sum_{l=-\infty}^{\infty}
 \frac{1}{(2\sqrt{\pi})^{D-1}}\frac{(\pi R)^{-D+1}}{\Gamma(\frac{D+1}{2})}
 \\ 
 &
 \int_0^\infty \rmd u
 (u^2+2\pi R|\omega|u)^{\frac{D-1}{2}}
 \frac{\rmd }{\rmd u}
 [
 \ln N(iu+i\pi R |\omega|)+\ln N(-iu-i\pi R |\omega|)]. 
\end{align}
In this method, the mass eigenvalues $m_n$ are not need to calculate the potential, 
 namely, the functions $N(z)$ of determining the mass spectrum are only required. 
Thus, the above formula allows us to calculate the contributions from the complicated KK mass spectrum.

To check the validity of the general potential derived above (\ref{eq:EPinfnitetemperature}), 
 we show an example of the calculation of the contributions from the $W$ boson 
 by the $\zeta$ function regularization method.
We note that the $W$ boson mass is given as a familiar form $m_n = \frac{n+\alpha}{R}$, 
 which satisfies the following relation
\begin{equation}
 \sin(\pi R m_n)-\sin\pi \alpha=0. 
\end{equation}
In this case, the function $N(z)$ that the KK mass spectrum satisfies is found as 
\begin{equation}
 N(z)=\sin z -\sin\pi \alpha. 
 \label{NZ}
\end{equation}
Substituting the $D=4$ and $N_\text{DOF}=3$,
we obtain
\begin{align}
 V_\text{eff}
 =&
 -\frac{3}{2\beta} \sum_{l=-\infty}^{\infty}
 \frac{1}{(2\sqrt{\pi})^{3}}\frac{(\pi R)^{-3}}{\Gamma(\frac{5}{2})}
 \int_0^\infty \rmd u
 (u^2+2\pi R|\omega|u)^{\frac{3}{2}} \times
 \nonumber \\
 &
 \frac{\rmd }{\rmd u}
 [
  \ln \{\sin(iu+i\pi R |\omega|)-\sin\pi\alpha\}+\ln \{\sin(-iu-i\pi R |\omega|)-\sin\pi\alpha\}]
  \\
 =&
 -\frac{1}{4\pi^5\beta R^3}
 \sum_{l=-\infty}^{\infty}
 \int_0^\infty \rmd u
 (u^2+2\pi R|\omega|u)^{\frac{3}{2}}
 \frac{\rmd }{\rmd u}
  \ln\frac{\cosh(2u+2\pi R |\omega|)+\cos 2\pi\alpha}{2}
\end{align}
where $\omega = \frac{2\pi l}{\beta}$.
Since this potential has a divergence independent of $\alpha$, 
 we subtract $V(\alpha=0)$ corresponding to the vacuum energy as the regularization.  
The effective potential at finite temperature is finally obtained as
\begin{align}
 \label{eq:integralformW}
  V_\text{eff}
 =&
 -\frac{1}{4\pi^5 \beta R^3}
 \sum_{l=-\infty}^\infty \int_0^\infty  \rmd u
 \left(u^2+4\pi^2{R}\frac{|l|}{\beta} u\right)^{3/2}
 \frac{\rmd}{\rmd u}\ln 
 \frac{\cosh(2u+4\pi^2{R}\frac{|l|}{\beta})-\cos(2\pi \alpha)}{\e^{2u+4\pi^2{R}\frac{|l|}{\beta}} }. 
\end{align}

Next, we calculate the contributions from the $W$ boson to 1-loop effective potential at finite temperature 
in terms of the Poisson resummation formula 
 and compare it with the above result calculated by the $\zeta$ function regularization method.
The result is given by
\begin{align}
 \label{eq:summationformW}
 V_\text{eff}^\text{Poisson}
 =
 -\frac{9}{2\pi}R
 \sum_{n=1}^ \infty 
 \frac{1}{(2\pi Rn)^{5}}
 \cos(2\pi n\alpha)
 -\frac{9}{\pi}R
 \sum_{n=1}^ \infty \sum_{l=1}^\infty
 \frac{1}{\{(2\pi Rn)^2+(\beta l)^2\}^{5/2}}
 \cos(2\pi n\alpha). 
\end{align}
The first and second term corresponds to the zero temperature 
 and the fintie temperature part of the effective potential, respectively.

The comparison between the results calculated in two different methods 
 is depicted in the Figure 2. 
In the left plot, the summation in both effective potentials 
 by $\zeta$ function regularization method and Poisson resummation is cut off 
 up to the finite number of modes. 
The plot expresses a dependence of the both potential on the maximum values. 
A good agreement between two potentials is seen for the maximum values more than $n=l=150$.
In the right plot, the dependences of the outline of the both potentials 
 on the various maximum values of the summation are shown. 
The integral in the potential calculated by $\zeta$ function regularization method converges more rapidly. 
From these comparisons, we confirm a validity of the 1-loop effective potential derived by $\zeta$ function regularization method.  
\begin{figure}
 \includegraphics[scale=0.4]{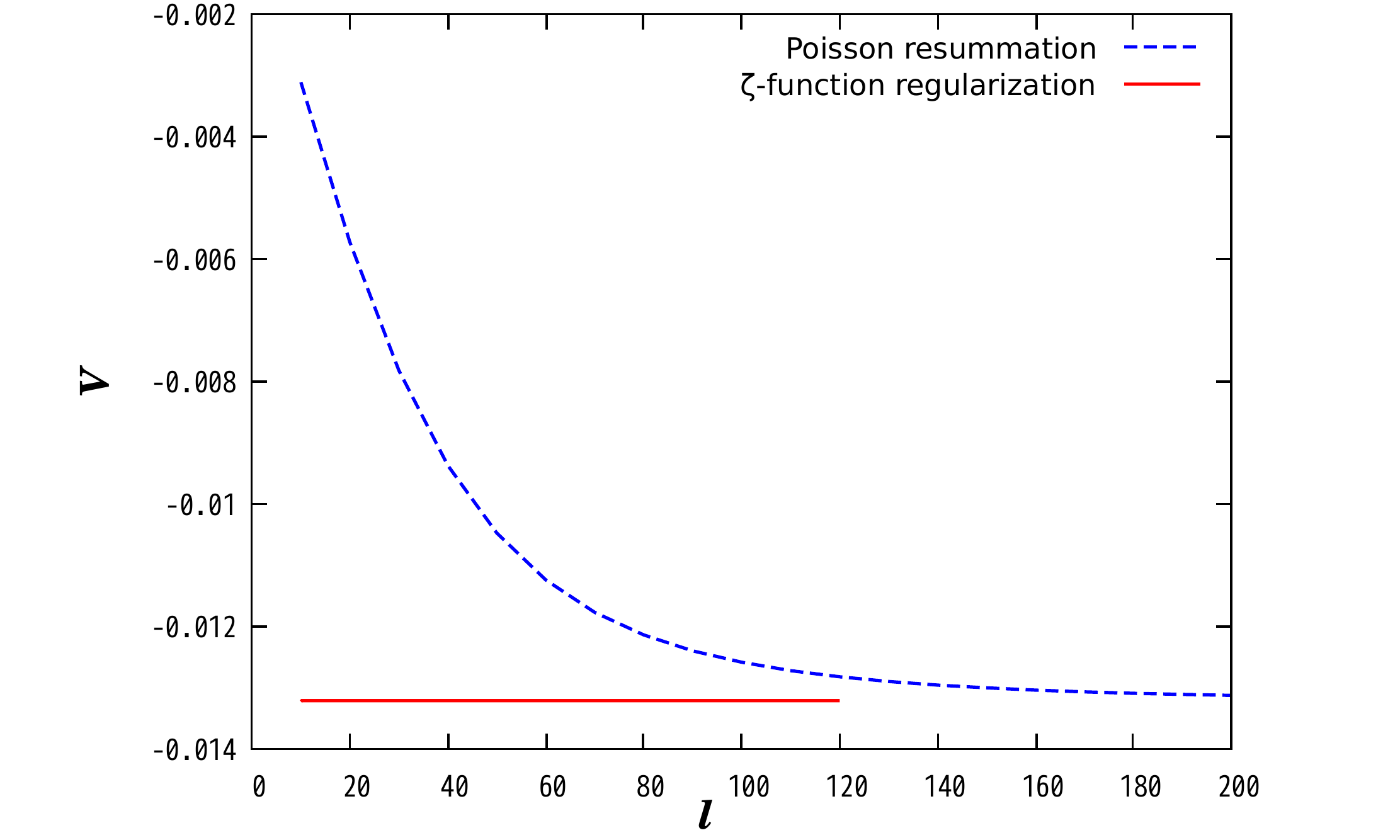}
 \includegraphics[scale=0.4]{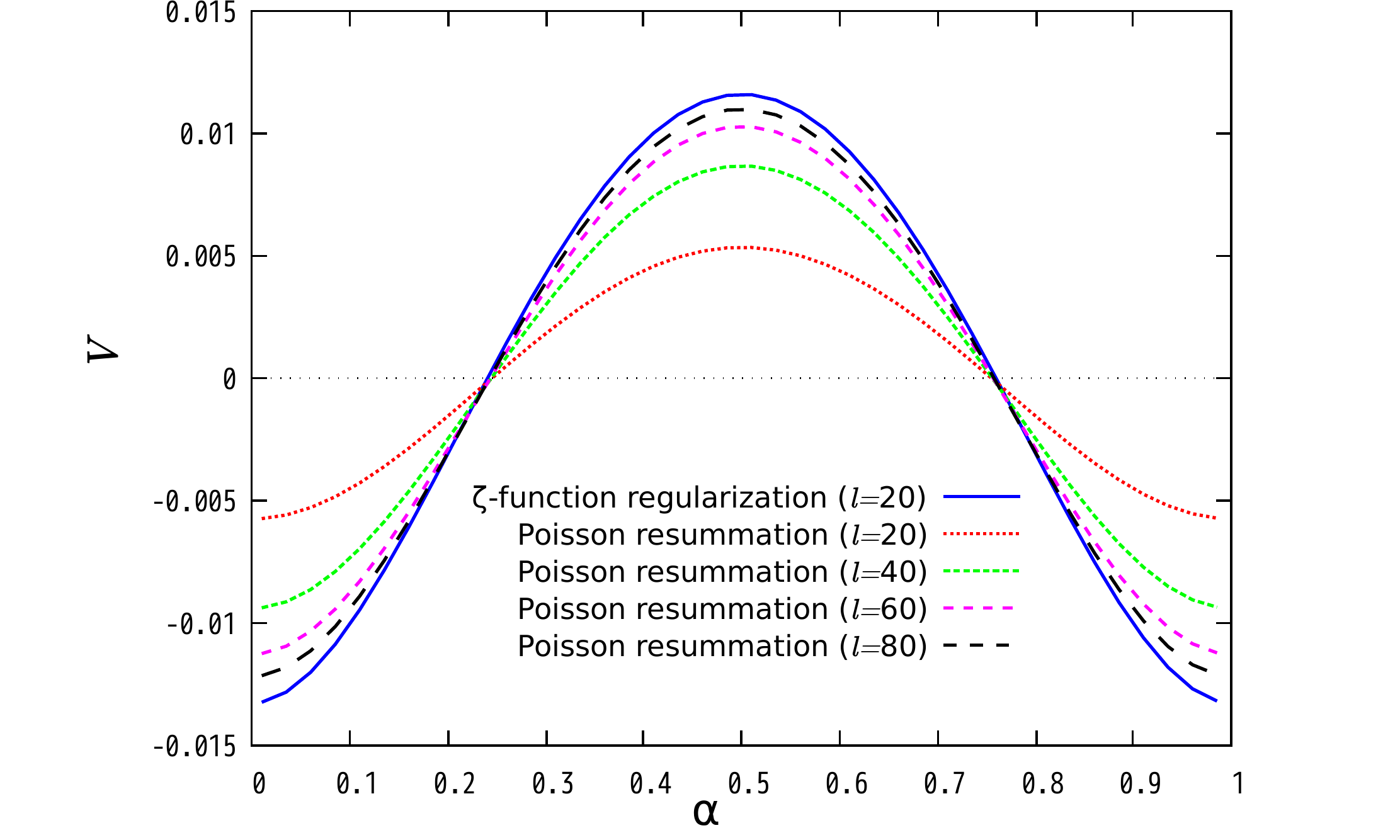}
 \label{figure:comparison}
 \caption{The comparison of convergence nature between two expressions 
eq.(\ref{eq:integralformW}) and eq.(\ref{eq:summationformW})
 The left figure shows the values of effective potential for $R^{-1}=1 \text{TeV}, \beta=0.1/\text{TeV}, \alpha=0.01$.  
 The horizontal axis describes an upper limit of summation ($n$ and $l$).
 The right one shows the effective potential in the various upper limit.
 }
\end{figure}

\section{Analysis of the electroweak phase transition at finite temperature}

Before studying the phase transition at finite temperature in our model, 
 we discuss general properties of the phase transition at finite temperature. 
The discussion in the case of GHU at finite temperature is found in \cite{MT, PS}. 
In order to realize the first order phase transition, 
 the cubic term in the 1-loop Higgs potential plays an essential role. 
The cubic terms arise only from the massless bosonic field contributions. 
In the case of higher dimensional gauge theories, 
 the phase transition is likely to be first order, 
 since the massless gauge boson contribution is present model independently. 
The discussions by use of approximate forms of the potential for the case of simple KK mass spectrum
 were given in \cite{MT, PS}. 
Also, this properties must be valid for the general potential derived in this paper.  
It is nontrivial and model dependent whether the phase transition is strong enough for the electroweak baryogenesis  
 and whether it is compatible with 125 GeV Higgs mass.

\begin{figure}
 \includegraphics[scale=0.4]{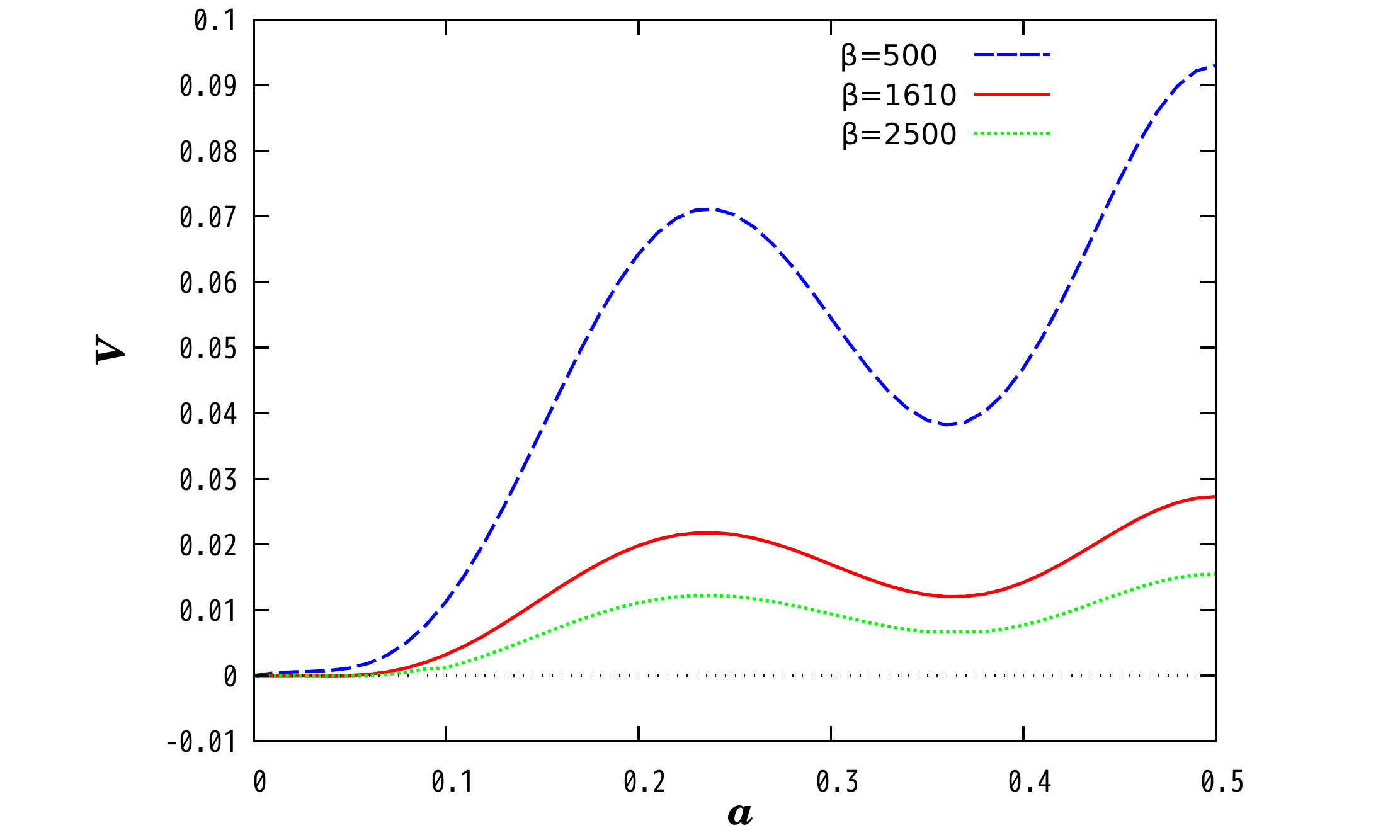}
 \includegraphics[scale=0.4]{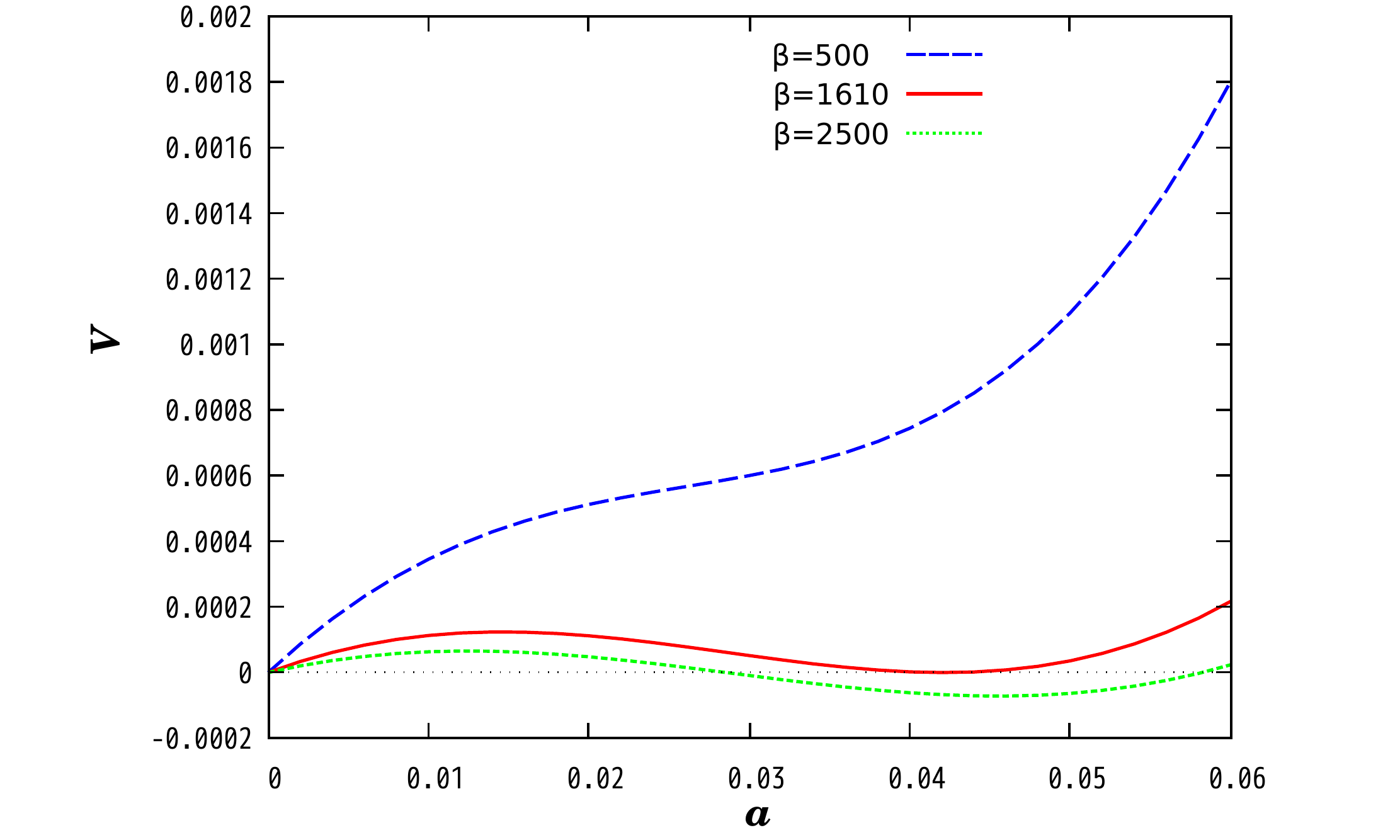}
 \caption{The effective potential at some particular temperatures. 
 Electroweak symmetry is restored at $\beta$=1610/TeV.}
 \label{fig:HP-all}
\end{figure}

Now, we are ready to discuss the electroweak phase transition at finite temperature in our GHU model. 
In order to compute Higgs potential of our model by using the result eq.~(\ref{eq:EPinfnitetemperature}),
We need seven kinds of functions, 
$N_Z,N_W,N_\text{BOT},N_\text{TOP},N_\text{LSM},N_\text{exotic}$ and $N_\text{M}$,
 which are functions to determine the mass spectrum 
 for the $Z$ and $W$ bosons, the bottom quark, top quark, 
 the SM fermions except for the top and bottom quarks, the exotic fermions and the mirror fermions. 
The explicit forms for the $W$ and $Z$ boson $N_{W,Z}$ were given in \cite{Adachi:2018mby} 
 and the other functions can be similarly obtained but are very complicated. 
Subtracting $V(\alpha =0)$ as was done in the above example, 
the four dimensional effective potential is obtained as
\begin{align}
V_\text{R}
 =& \nt
 -
 \frac{1}{64\pi^5 \beta R^3} \sum_{l=-\infty}^{\infty}
 \int_0^\infty \rmd u
 \\ \nt 
 &
 \Bigg[
 3(u^2+2\pi R|\omega_B|u)^{\frac{3}{2}}
 \frac{\rmd }{\rmd u}
 \Big\{
 \ln N_Z(iu+i\pi R |\omega_B|)+\ln N_Z(-iu-i\pi R |\omega_B|)
 \\ \nt &~~~~~~~~~~~~
+\ln N_W(iu+i\pi R |\omega_B|)+\ln N_W(-iu-i\pi R |\omega_B|)
\Big\}
\\ \nt
 &
 -
 3\cdot 4(u^2+2\pi R|\omega_F|u)^{\frac{3}{2}}
 \frac{\rmd }{\rmd u}
 \Big\{
 \ln N_\text{BOT}(iu+i\pi R |\omega_F|)+\ln N_\text{BOT}(-iu-i\pi R |\omega_F|)
 \\\nt &~~~~~~~~~~~~
 +\ln N_\text{TOP}(iu+i\pi R |\omega_F|)+\ln N_\text{TOP}(-iu-i\pi R |\omega_F|)
 \\\nt &~~~~~~~~~~~~
 +\ln N_\text{LSM}(iu+i\pi R |\omega_F|)+\ln N_\text{LSM}(-iu-i\pi R |\omega_F|)
 \\ \nt &~~~~~~~~~~~~
 +\ln N_\text{exotic}(iu+i\pi R |\omega_F|)+\ln N_\text{exotic}(-iu-i\pi R |\omega_F|)
 \\ &~~~~~~~~~~~~
 +\ln N_\text{M}(iu+i\pi R |\omega_F|)+\ln N_\text{M}(-iu-i\pi R |\omega_F|)
  \Big\}
  \Bigg] 
 -(\alpha\to 0). 
  \label{effpot}
\end{align}
Although their explicit expressions of the potential is omitted here since they are very lengthy and complicated, 
 they are written in our previous paper \cite{Adachi:2018mby}.  
We choose the compactification scale and the bulk mass for the third generation quarks 
 as $R^{-1}=1.43 {\rm TeV},M=1.95 {\rm TeV}$,
 which succeeds in explaining the quark mass parameters including top quark mass \cite{Adachi:2018mby}.
 The resultant figures are depicted in Fig. \ref{fig:HP-all}. 
The left plot shows the 1-loop Higgs effective potential at some temperatures. 
The temperature changes lower accordingly from the blue curve to the green one. 
The right plot zooms up around the minimum of the same potentials in the left plot.  
The potential minimum is at origin in the case of blue potential and the electroweak symmetry is unbroken. 
As the temperature is lowered, 
 the red potential minimum at origin and some finite VEV are degenerate at the critical temperature. 
Lowering the temperature further, 
 the green potential minimum is located at some VEV, which means that the electroweak symmetry is broken.  
 The critical temperature $T_{C}$ and the corresponding VEV at the critical temperature $\alpha(T_{C})$ 
  can be read off as $\beta_\text{C}= T_{C}^{-1}=1610/{\rm TeV}, \alpha(T_{C})=0.0422$. 
By using the relation $\dis v= \frac{\alpha}{Rg_4}$,
the ratio between the VEV at the critical temperature and the critical temperature, 
 which gives a signal of the first order phase transition, is found
\begin{equation}
 \frac{v(T_C)}{T_C}=
 \beta_C \frac{\alpha(T_C)}{R g_4}
 =1610\times \frac{0.0422}{1.43 g_4}
 =47.5 \frac{1}{g_4}, 
\end{equation}
where the 4D $SU(2)$ gauge coupling is smaller than the unity $g_4 < 1$, 
 which ensures the strong first order phase transition in our model.
This result opens a possibility to apply to the electroweak baryogenesis 
 since the third one of Sakharov's three conditions required for the baryon asymmetry is satisfied. 

Such a relatively lower critical temperature compared with the compactification scale is the general feature in our model.
As we mentioned in the section \ref{section:model}, 
the small VEV compared with the compactification scale is achieved 
 by introducing the higher rank representations such as the $\bf 15$ represenations.
It indicates that the magnitude of the minimum of the effective potential is much smaller than the model such as \cite{PS},
 so that the electroweak symmetry is restored at low temperature in our model.

\section{Summary}
In this paper, we have studied the electroweak phase transition at finite temperature 
 in a model proposed by the authors, 5D $SU(3) \otimes U(1)_X$ GHU 
 with a realistic fermion mass hierarchy including top quark mass, 
 a successful electroweak symmetry breaking 
 and an observed Higgs boson mass \cite{Adachi:2018mby}.  
The purpose of the analysis is to investigate whether the phase transition is strong first order or not.   
As an application of the electroweak phase transition, the electroweak baryogenesis is very familiar. 
In order to work the mechanism, the famous Sakharov's conditions must be satisfied. 
One of the conditions is that the phase transition should be strong first order. 

In order to study the phase transition in our model, 
 the calculation of the 1-loop effective potential is not trivial 
 since our model has the brane localized fermions and their couplings to bulk fermions 
 and the KK mass spectrum of such fermions cannot be exactly found in general. 
Therefore, we cannot use Poisson resummation formula in calculation of the 1-loop effective potential. 
Instead, we have derived a general formula of the 1-loop effective potential at finite temperature 
 by employing $\zeta$ function regularization method. 
The advantage of this method is that even if the KK mass spectrum cannot be found, 
 the 1-loop effective potential can be calculated if we have the functions determining the KK mass spectrum. 
Remarkably, the general formula of the 1-loop effective potential at finite temperature 
 can be applicable to any higher dimensional theory in flat space 
 with one extra spatial dimension compactified. 

Applying the general potential to our GHU model, 
 we have analyzed the electroweak phase transition at finite temperature. 
The strong first order phase transition was found at relatively low temperature 
 and the 125 GeV Higgs mass was also compatible. 
The results are a great step towards the realization of the electroweak baryogenesis in GHU.  
There are still two Sakharov's conditions to be satisfied, the baryon number violation and CP, C violations.  
CP violation is most nontrivial in GHU 
 since Yukawa coupling is originated from the gauge coupling. 
Therefore, Yukawa couplings are real as they stands. 
Furthermore, it is also nontrivial to obtain an enough CP phase required for the electroweak baryogenesis, 
which must be an additional CP phase other than Kobayahi-Maskawa CP phase. 
Our previous work on CP violation in the context of GHU \cite{CP} might help to overcome this issue.

As another application, it would be interesting to investigate a tension 
 between the first order phase transition and the deviation from the SM prediction of the Higgs self coupling \cite{GSW, KOS}. 
 In our previous work on the triple Higgs coupling in GHU \cite{AMTHC}, 
  a sizable deviation (around 70\%) from the SM prediction has been obtained. 
Combining analysis of this paper and \cite{AMTHC} will be left for our future work. 

\section*{Acknowledgments}
The work of N.M. is supported in part by JSPS KAKENHI Grant Number JP17K05420.


\appendix

\section{$\zeta$ function regularization method}
In this appendix, 
we derive the 1-loop effective potential at finite temperature 
 by utilizing the $\zeta$ function regularization method.
The Poisson resummation formula is an easy way to calculate the effective potential 
 in the case for the simple KK mass spectrum,
 but it is very difficult in case that the KK mass spectrum cannot be solved exactly or approximately.
It is because the explicit expression of mass eigenvalues are needed to carry out the KK mode summation. 
Such a complicated mass spectrum often appears in models with the brane-localized terms as in our model.
In the $\zeta$ function regularization method,
the KK mode summation is replaced with a contour integration thanks to the Cauchy's theorem 
 and the functions determining the KK mass spectrum instead of the KK mass spectrum themselves 
 are necessary for the integration.
It allows us to calculate the 1-loop effective potential 
 without the explicit solutions of KK mass spectrum.


\subsection{Formula}
In this subsection,
  we explain a derivation of the general formula for the 1-loop effective potential at finite temperature in detail. 
We begin with the contributions from the bosonic/fermionic field in $D+1$ dimensional space-time 
 where only one spatial extra dimension is compactified on $S^1/Z_2$.
The formula is given by
\begin{equation}
V
 =(-1)^F
 \frac{N_\text{DOF}}{\beta} \sum_{l=-\infty}^{\infty}\sum_{n=-\infty}^{\infty}
 \int\frac{\rmd^{D-1} p}{(2\pi)^{D-1}}\frac12 \ln(\vec p^2 +\omega^2 +m_n^2)
\end{equation}
 where $\beta=T^{-1}$ denotes an inverse temperature. 
The $\omega$ stands for the Matsubara frequency, 
 in which $\omega_B=\frac{2\pi}{\beta}l$ is for the bosonic field and 
$\omega_F=\frac{\pi}{\beta}(2l-1)$ is for the fermionic field.  
The $N_\text{DOF}$ stands for the degrees of freedom of the field running in the loop.
$F$ means a fermion number. 
Note that the momentum integral is a $D-1$ dimensional 
 since the time component is compactified on a circle 
 in the imaginary time formalism of the finite temperature theory.
The above 1-loop effective potential can be rewritten by the $\zeta$ function method as follows
\begin{align}
V
 =& 
 -(-1)^F
 \frac{N_\text{DOF}}{2\beta} \sum_{l=-\infty}^{\infty}
 \frac{1}{(2\sqrt{\pi})^{D-1}}\frac{(\pi R)^{-D+1}}{\Gamma(\frac{D+1}{2})} \times
 \nt \\ 
 &
 \int_0^\infty \rmd u
 (u^2+2\pi R|\omega|u)^{\frac{D-1}{2}}
 \frac{\rmd }{\rmd u}
 [
 \ln N(iu+i\pi R |\omega|)+\ln N(-iu-i\pi R |\omega|)]
\end{align}
where $\Gamma(x)$ is the Gamma function 
and the function $N(z)$ satisfies the $N(\pi R m_n)=0$ for the KK mass spectrum $m_n$. 

\begin{figure}[h]
 \begin{center}
  \includegraphics{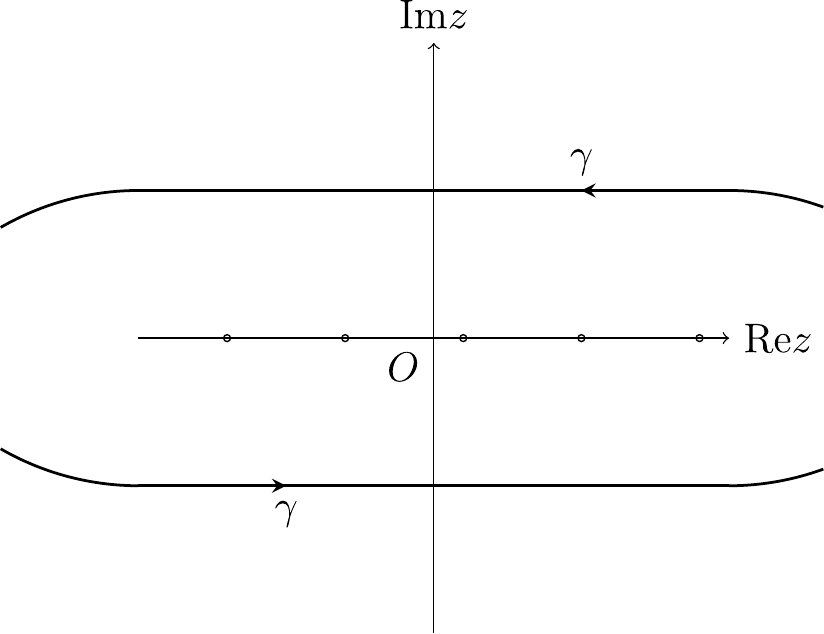}
 \end{center}
\caption{Contour: Small circles describe poles  on the real axis.}
 \label{figure:contourpath} 
\end{figure}

\begin{figure}[h]
\begin{center}
 \includegraphics{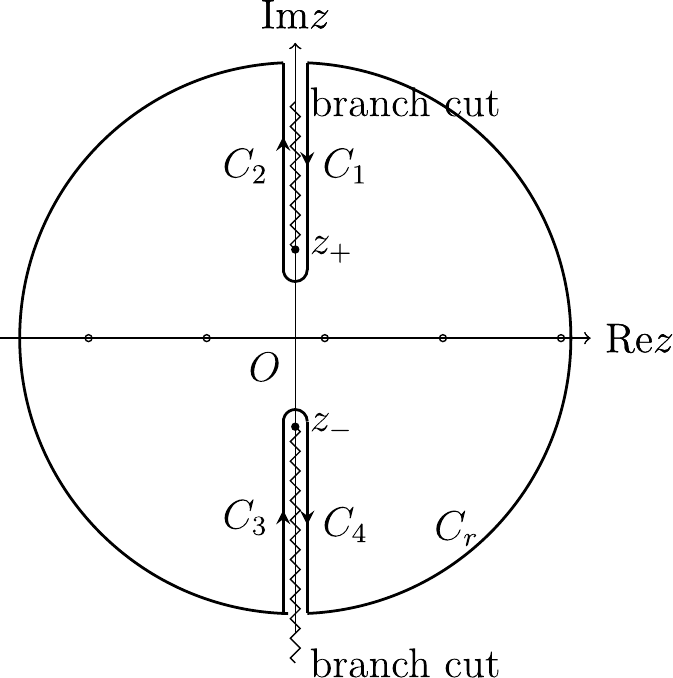}
\end{center}
\caption{Deformed contour: The zigzag line shows branch cuts.}
 \label{figure:contournonzero}
\end{figure}

In order to derive the above result, 
 we first apply the following relation to
 carry out the momentum integration.
\begin{equation}
 \ln (\vec p^2 + \omega ^2 +m_n^2)
 =-\partial_s  (\vec p^2 +\omega^2 +m_n^2) ^{-s} |_{s=0} 
 =\left.
 -\partial_s \frac{1}{\Gamma(s)}\int_0^\infty t^{s-1}
 \e^{-(\vec p^2 +\omega^2 +m_n^2)t}\right|_{s=0}.
\end{equation}
After integrating out the momentum $\vec p$,
we have 
\begin{align}
V
 =&
 \left.
 (-1)^F\frac{N_\text{DOF}}{2\beta} 
 \sum_{l=-\infty}^{\infty}\sum_{n=-\infty}^{\infty}
 \frac{-\partial_s}{(2\sqrt{\pi})^{D-1}}
 \frac{\Gamma(s-\frac{D-1}{2})}{\Gamma(s)}
 (m_n^2+\omega^2)^{-s+\frac{D-1}{2}}
 \right |_{s=0}. 
\end{align}
It is easy to sum up with respect to the index $n$ 
 in the case for the simple mass spectrum such as $m_n=n/R$,
 however, we will consider the case that the exact mass spectrum cannot be solved.
The mass spectrum $m_n$ satisfies $N(\pi Rm_n)=0$ through some function $N(z)$ 
 which can be obtained from the boundary conditions at orbifold fixed points,
 and the mode summation can be replaced with the contour integral as
\begin{equation}
 \sum_n (m_n^2+\omega^2)^{-s + \frac{D-1}{2}}  
 =(\pi R)^{2s - D-1}\int_\gamma \frac{\rmd z}{2\pi i}(z^2+\pi^2 R^2\omega^2)^{-s + \frac{D-1}{2}}
 \frac{\rmd }{\rmd z} N(z).
\end{equation}
Note that the contour $\gamma$ is chosen to surround the real axis
 in the counterclockwise direction as shown in Fig. \ref{figure:contourpath}.
Thus we have 
\begin{align}
 V&=(-1)^F\frac{N_\text{DOF}}{2\beta} \sum_{l=-\infty}^{\infty}
 \frac{-\partial_s}{(2\sqrt{\pi})^{D-1}}\frac{\Gamma(s-\frac{D-1}{2})}{\Gamma(s)} \times \nonumber \\
& \hspace*{10mm} (\pi R)^{-D+1+2s}
 \int_\gamma \frac{\rmd z}{2\pi i}(z^2+\pi^2R^2\omega^2)^{-s+\frac{D-1}{2}}
 \frac{\rmd }{\rmd z}\ln (N(z))|_{s=0}.
\end{align}
 To evaluate the integration, we push up the contour $\gamma$.
 In deforming the contour, the path should not across the branch cut
 since the integrand is a multi-valued function.
We introduce $p$ and $z_{\pm}$ as $p=-s+\frac{D-1}{2}$ 
 and $z_\pm = \e^{\pm \frac{i}{2}\pi}\pi R\omega_l$ to simplify the expression.
 The integrand is factored out as
  \begin{equation}
   (z^2+\pi^2R^2\omega^2)^{-s+\frac{D-1}{2}}
   =(z-z_+)^p(z-z_-)^p. 
  \end{equation}
As mentioned the above, since the integrand is a multi-valued function, 
 we need the branch cuts along with the imaginary axis as shown in Fig. \ref{figure:contournonzero}.
The $C_1,C_2,C_3$ and $C_4$ are the paths which are deformed from the contour $\gamma$.
They are parameterized by pure imaginary,
 however, the integrand shifts the phases when the path changes from $C_1$ to $C_2$,
or from $C_3$ to $C_4$.

Now we evaluate the contributions from the paths $C_1$ and $C_2$.
We parametrize the variables $u$ 
 as $z=z_++\e^{\frac{1}{2}i \pi}u$ on the path $C_1$ 
 and $z=z_++\e^{-\frac{3}{2}i\pi}u$ on the path $C_2$. 
Turning around the branch point $z=z_+$ in the counterclockwise direction, 
 the term $(z-z_+)^p$ in the integrand shifts the phase $\e^{-2\pi ip}$.
Note that the other term $(z-z_-)^p$ in the integrand does not shift the phase 
 because it is continuous in $\text{Im}(z)>0$ region.
Thus the contributions from the paths $C_1$ and $C_2$ are calculated as
\begin{align}
 &\int_{C_1} \frac{\rmd z}{2\pi i}
 (z-z_+)^p(z-z_-)^p\frac{\rmd}{\rmd z}\ln N(z)
 +
 \int_{C_2} \frac{\rmd z}{2\pi i}
 (z-z_+)^p(z-z_-)^p\frac{\rmd}{\rmd z}\ln N(z)   
  \nt \\
 =&
 \int_\infty^0 \frac{i\rmd u}{2\pi i}
 (\e^{\frac{i}{2}\pi}u)^p(iu+z_+-z_-)^p
 \frac{\rmd}{i\rmd u}\ln N(iu +z_+)
 \nt \\
 &+
 \int^\infty_0 \frac{i\rmd u}{2\pi i}
 (\e^{-\frac{3}{2}i\pi}u)^p(iu+z_+-z_-)^p
 \frac{\rmd}{i\rmd u}\ln N(iu +z_+)
 \nt \\
 =&
 \int^\infty_0 \frac{\rmd u}{2\pi i}
 (-\e^{\frac{i}{2}\pi p}+\e^{-\frac{3}{2}i\pi p})u^p\e^{\frac{i}{2}\pi p}
 \left[u+\frac{2}{i}z_+\right]^p
 \frac{\rmd}{\rmd u}\ln N(iu+z_+)
 \nt \\
 =&
 -\frac{\sin\pi  p}{\pi}\int^\infty_0 \rmd u
 u^p(u+2R\omega_l)^p
 \frac{\rmd}{\rmd u}\ln N(iu+z_+). 
\end{align}
The same procedure can be applied to the path $C_3$ and $C_4$.
We parametrize the variable $u$ as $z=z_-+\e^{\frac{3}{2}\pi}u$ on the path $C_3$ 
 and $z=z_++\e^{-\frac{i}{2}\pi}u$ on the path $C_4$.
In this case,
 the term $(z-z_-)^p$ in the integrand shifts the phases $\e^{-2\pi p}$ at the branch point $z=z_-$.
The other term $(z-z_+)^p$ in the integrand is continuous in $\text{Im}(z)<0$ region 
 and no phase shifts is present. 
The contributions from the paths $C_3$ and $C_4$ are similarly obtained
\begin{align}
  &
 \int_{C_3} \frac{\rmd z}{2\pi i}
 (z-z_+)^p(z-z_-)^p\frac{\rmd}{\rmd z}\ln N(z)
 +
 \int_{C_4} \frac{\rmd z}{2\pi i}
 (z-z_+)^p(z-z_-)^p\frac{\rmd}{\rmd z}\ln N(z)
 \nt \\
 =&
 -\frac{\sin\pi p}{\pi}
 \int_0^\infty\rmd u (u^2+2R\omega_l u)^p
 \frac{\rmd}{\rmd u}\ln N(-iu+z_+). 
\end{align}
To summarize, the contour integration becomes  
\begin{align}
 V
 =&
 -(-1)^F\frac{N_\text{DOF}}{2\beta} \sum_{l=-\infty}^{\infty}
 \frac{-\partial_s}{(2\sqrt{\pi})^{D-1}}
 \frac{\Gamma(s-\frac{D-1}{2})}{\Gamma(s)}(\pi R)^{-D+1+2s}
 \frac{\sin\pi(\frac{D-1}{2}-s)}{\pi}
 \nt \\
 &
  \times \int_0^\infty \rmd u
 (u^2+2\pi R|\omega_B|u)^{\frac{D-1}{2}-s}
 \frac{\rmd }{\rmd u}
 [\ln N(iu+z_+)+\ln N(-iu+z_-)]
 _{s=0}. 
\end{align}
Before carrying out the $s$ derivative, it is very useful to rewrite the following term 
 by applying the Euler's reflection formula, 
\begin{align}
 \frac{\Gamma(s-\frac{D-1}{2})}{\Gamma(s)}\frac{\sin\pi p}{\pi}
 =&
 \frac{\Gamma(s-\frac{D-1}{2})}{\Gamma(s)}\frac{1}{\Gamma(p)\Gamma(1-p)}
 \nt \\
 =&
 \frac{1}{s-\frac{D-1}{2}}\frac{\sin \pi s}{\pi}\Gamma(1-s)
 \frac{1}{\Gamma(\frac{D-1}{2}-s)}. 
\end{align}
Since $\sin \pi s=0$ at $s=0$, 
the potential is nonvanishing only when the $\partial_s$ operates on the $\sin\pi s$.
Then we arrive at the final result
\begin{align}
 V
 =&
 -(-1)^F
 \frac{N_\text{DOF}}{2\beta} \sum_{l=-\infty}^{\infty}
 \frac{1}{(2\sqrt{\pi})^{D-1}}\frac{(\pi R)^{-D+1}}{\Gamma(\frac{D+1}{2})} \times
 \nt \\
 &
 \int_0^\infty \rmd u
 (u^2+2\pi R|\omega|u)^{\frac{D-1}{2}}
 \frac{\rmd }{\rmd u}
 [
 \ln N(iu+i\pi R |\omega|)+\ln N(-iu-i\pi R |\omega|)]. 
\end{align}





\end{document}